\documentclass[epj,twocolumn]{webofc}
\usepackage[varg]{txfonts}   
\def\nonum\\{\nonumber\\}
\def\bs#1{\boldsymbol{#1}}

\def\zI{\mathrm{i}\hspace{0.2mm}}
\woctitle{FUSION14}
\begin{document}
\title{Time-dependent Hartree-Fock calculations for multi-nucleon transfer processes}
%
\subtitle{Effects of particle evaporation on production cross sections}
\author{%
Kazuyuki Sekizawa\inst{1}\fnsep\thanks{\email{sekizawa@nucl.ph.tsukuba.ac.jp}}
\and
Kazuhiro Yabana\inst{1,2}\fnsep\thanks{\email{yabana@nucl.ph.tsukuba.ac.jp}}
}
\institute{%
Graduate School of Pure and Applied Sciences, University of Tsukuba, Tsukuba 305-8571, Japan
\and
Center for Computational Sciences, University of Tsukuba, Tsukuba 305-8577, Japan
}
\abstract{%
We present a microscopic calculation of multi-nucleon transfer
reactions employing the time-dependent Hartree-Fock (TDHF) theory.
In our previous publication [Phys. Rev. C \textbf{88}, 014614 (2013)], 
we reported our analysis for the multi-nucleon transfer processes for
several systems. Here we discuss effects of particle evaporation processes
on the production cross sections. Since particle evaporation processes
may not be described adequately by the TDHF calculations, we evaluate
them using a statistical model. As an input of the statistical model,
excitation energies of the final fragments are necessary. We evaluate
them from the TDHF wave function after collisions, extending the
particle number projection technique. From the calculation, the
particle evaporation effects are found to improve descriptions of
the production cross sections. However, the production cross sections
are still underestimated for processes where a number of protons are
transferred. Possible origins of the discrepancy are discussed.
}
\maketitle
%

\section{Introduction}\label{intro}

In low-energy nuclear reactions at around the Coulomb barrier,
multi-nucleon transfer (MNT) reactions are commonly observed
at an impact parameter region slightly outside that corresponding
to fusion reactions. The MNT reaction at around the Coulomb barrier
contains rich physics related to both structural properties and
time-dependent dynamics of colliding nuclei, and has been attracting
much interests in a number of aspects. Properties of a neck formed
at the distance of closest approach reflect structural and excitation
properties of the colliding nuclei \cite{Simenel(CC)}. Exchanges
of a number of nucleons caused by the neck formation can induce 
an energy dissipation, a transfer of energy from translational relative
motion to that of internal excitations. This provides an opportunity to
study correlation effects in nuclear dynamics \cite{Simenel(EPJ)}.
MNT reactions have also been expected to be a new means to 
produce unstable nuclei whose production is difficult by other 
methods \cite{Zagrebaev,KISS,Simenel(review)}. 

To uncover microscopic reaction mechanisms of MNT reactions 
and to predict preferable conditions to produce objective nuclei, 
we have undertaken microscopic calculations of the MNT reactions 
employing the time-dependent Hartree-Fock (TDHF) theory.
We presented results of the TDHF calculations for MNT processes
in $^{40,48}$Ca+$^{124}$Sn, $^{40}$Ca+$^{208}$Pb,
and $^{58}$Ni+$^{208}$Pb reactions, for which extensive
measurements are available \cite{KS_KY_MNT}. To calculate
transfer probabilities and cross sections, we used a particle
number projection technique which was recently proposed by
C.~Simenel \cite{Projection}. From the results, we found that
the TDHF calculations can reproduce measured cross sections
quantitatively, when the number of transferred nucleons is small.
However, as the number of transferred nucleons increases,
a peak position of the calculated cross sections shifts towards
larger neutron and proton number sides compared with
measurements. One of possible origins of the discrepancy 
may be an insufficient description of particle evaporation 
processes in our TDHF calculations. Because we calculated the
transfer probabilities and the cross sections from a TDHF wave
function just after two nuclei separates (typically, order of
$10^{-21}$~s after the reseparation), effects of secondary
evaporation processes which occur in much longer time scale
have not been taken into account in the calculated cross sections.

In the present article, we will show an outline of our recent
attempt to evaluate effects of particle evaporation processes on
the production cross sections based on the TDHF theory \cite{article2}.
To estimate how many nucleons are to be evaporated from the
produced fragment, we need to calculate excitation energy
of the fragment in each transfer channel. We calculate the
excitation energy extending the particle number projection
technique which was originally used to calculate transfer 
probabilities from a TDHF wave function after collision. 
We then evaluate evaporation probabilities by employing a 
statistical model \cite{DostrovskyIII} in which the excitation 
energy obtained from the projection analysis will be used
as an input. We finally calculate the production cross sections
including evaporation effects.

It is certainly true that correlation effects included in the
TDHF theory is rather limited. For example, isoscalar and
isovector pair transfers, $\alpha$-cluster transfer are not
treated adequately. We consider that MNT cross sections with
improved treatment of evaporation processes will help to
uncover what is lacking and what is needed in the TDHF
calculation for the MNT processes. One of the aims of this
work is to clarify to what extent the TDHF theory can describe
MNT cross sections quantitatively if we include the effects 
of particle evaporation processes. 

This article is organized as follows. In Sec.~\ref{Sec:formulation},
we present an outline of our formalism to include the effect of
the particle evaporation processes in the calculation of the production
cross sections. In Sec.~\ref{Sec:results}, we show calculated
production cross sections for $^{58}$Ni+$^{208}$Pb reaction
with and without particle evaporation effects. In Sec.~\ref{Sec:summary},
a summary and a future prospect are presented.

\section{Formulation}\label{Sec:formulation}

\subsection{Particle number projection technique}

In the TDHF theory, a many-body wave function of the system
is described by a single Slater determinant composed of
single-particle wave functions of nucleons. We denote the 
number of neutrons and protons in the projectile (target)
as $N_\mathrm{P}^{(n)}$ ($N_\mathrm{T}^{(n)}$)
and $N_\mathrm{P}^{(p)}$ ($N_\mathrm{T}^{(p)}$), respectively. 
The total numbers of neutrons and protons are then given by $N^{(n)}=
N_\mathrm{P}^{(n)}+N_\mathrm{T}^{(n)}$ and $N^{(p)}=
N_\mathrm{P}^{(p)}+N_\mathrm{T}^{(p)}$, respectively.
We denote the total number of nucleons as $A=N^{(n)}+N^{(p)}$.
In the TDHF theory, the many-body wave function of the system
is expressed as a direct product of Slater determinants for neutrons
and protons,
\begin{equation}
\Psi(x_1,\cdots,x_A) =
\Psi^{(n)}(x_1,\cdots,x_{N^{(n)}}) \otimes
\Psi^{(p)}(x_1,\cdots,x_{N^{(p)}}),
\end{equation}
where
\begin{equation}
\Psi^{(q)}(x_1,\cdots,x_{N^{(q)}}) =
\frac{1}{\sqrt{N^{(q)}!}}\det\bigl\{ \psi_i^{(q)}(x_j) \bigr\}
\end{equation}
denotes the many-body wave function for neutrons ($q=n$)
or protons ($q=p$). $\psi_i^{(q)}(x)$ is the single-particle
wave function of a nucleon having an isospin $q$, where
$x$ denotes a set of spatial and spin coordinates,
$x \equiv ({\bf r},\sigma)$.

During the collision, single-particle wave functions 
which originally belong to either projectile or target
region extend to a whole spatial region where a mean-field 
potential of the collision partner exists.
After the collision, the whole system separates into
two fragments, a projectile-like fragment (PLF) and 
a target-like fragment (TLF). We divide the whole space into 
two regions, a projectile region $V_\mathrm{P}$ which 
includes the PLF and a target region $V_\mathrm{T}$ 
which includes the TLF. Each single-particle wave function
extends in both $V_\mathrm{P}$ and $V_\mathrm{T}$.
Then, the TDHF wave function after collision is not an 
eigenstate of the number operator in respective regions, 
$\hat{N}_\mathrm{P}^{(q)}$ and $\hat{N}_\mathrm{T}^{(q)}$ 
($q=n$ for neutrons, $q=p$ for protons), but a superposition 
of states with different particle number distributions.

To calculate the particle number distributions, a particle number
projection technique was recently proposed by C.~Simenel
\cite{Projection}. A particle number projection operator which
projects out a particle number eigenstate of the number operators,
$\hat{N}_\mathrm{P}^{(q)}$ and $\hat{N}_\mathrm{T}^{(q)}$,
with eigenvalues, $n$ and $N^{(q)}-n$, respectively, can be expressed as
\begin{equation}
\hat{P}_n^{(q)} =
\frac{1}{2\pi} \int_0^{2\pi} \hspace{-1mm}d\theta\;
e^{\zI (n-\hat{N}_\mathrm{P}^{(q)})\theta}.
\end{equation}
The probability to find a nucleus composed of $n$ neutrons and
$z$ protons in the spatial region $V_\mathrm{P}$ is given by
\begin{equation}
P_{n,z} = P_n^{(n)} P_z^{(p)},
\label{P_nz}
\end{equation}
where
\begin{equation}
P_n^{(q)} \equiv \bigl<\Psi^{(q)}\big|\hat{P}_n^{(q)}\big|\Psi^{(q)}\bigr>
= \frac{1}{2\pi} \int_0^{2\pi} \hspace{-1mm} d\theta\;
e^{\zI n\theta} \det\mathcal{B}^{(q)}(\theta)
\end{equation}
and
\begin{equation}
\Bigl( \mathcal{B}^{(q)}(\theta) \Bigr)_{ij} \equiv
\bigl<\psi_i^{(q)}\big|\psi_j^{(q)}\bigr>_{V_\mathrm{T}}+
e^{-\zI\theta}\bigl<\psi_i^{(q)}\big|\psi_j^{(q)}\bigr>_{V_\mathrm{P}}.
\end{equation}
We have introduced a shorthand notation, $\bigl<\psi_i^{(q)}\big|\psi_j^{(q)}
\bigr>_\tau \equiv \sum_\sigma \int_\tau d{\bf r}\, \psi_i^{(q)*}({\bf r},\sigma)
\psi_j^{(q)}({\bf r},\sigma)$ ($\tau=V_\mathrm{P}$ or $V_\mathrm{T}$).
Because the TDHF wave function is a direct product of wave
functions for neutrons and protons, the probability is also
given by a product of probabilities for neutrons and protons
as Eq.~(\ref{P_nz}).

\subsection{Excitation energies of produced fragments}

We have extended the particle number projection technique 
to calculate an expectation value of an arbitrary operator,
using the particle number projected wave function \cite{article2}. 
The energy expectation value of the PLF composed of $n$ neutrons 
and $z$ protons is given by
\begin{equation}
E_{n,z}^\mathrm{PLF} \equiv
\frac{\bigl<\Psi\big|\hat{H}_{V_\mathrm{P}}
\hat{P}_n^{(n)}\hat{P}_z^{(p)}\big|\Psi\bigr>}
{\bigl<\Psi\big|\hat{P}_n^{(n)}\hat{P}_z^{(p)}\big|\Psi\bigr>}.
\label{def_E_nz}
\end{equation}
$\hat{H}_{V_\mathrm{P}} \equiv
\hat{T}_{V_\mathrm{P}}+\hat{V}_{V_\mathrm{P}}=
\sum_i \Theta_{V_\mathrm{P}}({\bf r}_i) \,\hat{t}_i+
\sum_{i<j}\Theta_{V_\mathrm{P}}({\bf r}_i)
\Theta_{V_\mathrm{P}}({\bf r}_j) \,\hat{v}_{ij}$
denotes a Hamiltonian acting only for the PLF, where we
have introduced a space division function,
$\Theta_{V_\mathrm{P}}({\bf r}) \equiv 1$ for ${\bf r}
\in V_\mathrm{P}$ and 0 for ${\bf r} \notin V_\mathrm{P}$.

In practice, we need to remove an energy associated with
the center-of-mass motion of the fragment.
For this purpose, we move to the rest frame of the PLF
before calculating the energy expectation value.
Denoting the mass and coordinates of the PLF
at the final stage of the calculation as $M_\mathrm{PLF}$ and
${\bf R}_\mathrm{PLF}(t_f)$, respectively, the wave vector of
the fragment is evaluated as ${\bf K}_\mathrm{PLF}=
M_\mathrm{PLF}\dot{\bf R}_\mathrm{PLF}(t_f)/\hbar$, where
$\dot{\bf R}_\mathrm{PLF}(t_f) \equiv \bigl[{\bf R}_\mathrm{PLF}
(t_f+\Delta t)-{\bf R}_\mathrm{PLF}(t_f-\Delta t)\bigr]/(2\Delta t)$.
We multiply the plane wave $e^{-\zI{\bf K}_\mathrm{PLF}
\bs{\cdot}{\bf r}}$ to the wave function in the spatial region
$V_\mathrm{P}$. 

After the removal of the center-of-mass motion of the fragment,
we evaluate the energy expectation value using Eq.~(\ref{def_E_nz}). 
The kinetic energy term for the PLF composed of $n$ neutrons and
$z$ protons can be calculated as
\begin{equation}
\mathcal{E}_{n,z,\mathrm{kin}}^\mathrm{PLF} =
\mathcal{E}_{n,\mathrm{kin}}^{(n)\,\mathrm{PLF}} +
\mathcal{E}_{z,\mathrm{kin}}^{(p)\,\mathrm{PLF}},
\end{equation}
where
\begin{eqnarray}
\mathcal{E}_{n,\mathrm{kin}}^{(q)\,\mathrm{PLF}} & \equiv &
\frac{1}{2\pi P_n^{(q)}}
\int_0^{2\pi} \hspace{-1.5mm} d\theta \;
e^{\zI n\theta} \det\mathcal{B}^{(q)}(\theta) \nonum\\[0.5mm]
&& \hspace{-21mm} \times \; \frac{\hbar^2}{2m} \,
\sum_{i=1}^{N^{(q)}}\sum_\sigma \,\int_{V_\mathrm{P}} d{\bf r} \;
\bs{\nabla}\psi_i^{(q)*}({\bf r},\sigma) \bs{\cdot}
\bs{\nabla}\tilde{\psi}_i^{(q)}({\bf r},\sigma,\theta).
\label{def_E_nz_kin}
\end{eqnarray}
$\tilde{\psi}_i^{(q)}({\bf r},\sigma,\theta)$
is defined by $\sum_{k \in q}
\Bigl( \mathcal{B}^{(q)}(\theta) \Bigr)^{-1}_{ik} \psi_k^{(q)}
({\bf r},\sigma)$.
The center-of-mass correction is simply taken into account
by considering the one-body term, replacing the coefficient
of the kinetic energy operator $\frac{\hbar^2}{2m}$ with
$\frac{\hbar^2}{2m}(1-\frac{1}{a})$, where $a=n+z$
denotes the mass number of the PLF. The interaction part is
calculated using transition densities, (e.g. the transition proton
density is given by $\tilde{\rho}^{(p)}({\bf r},\theta)
\equiv \sum_{i \in p, \sigma}\psi_i^*({\bf r},\sigma)
\tilde{\psi}_i({\bf r},\sigma,\theta)$). The two-body and
three-body interaction terms for the fragment composed of
$n$ neutrons and $z$ protons are calculated as
\begin{eqnarray}
\mathcal{E}_{n,z,\mathrm{int}}^\mathrm{PLF} & \equiv &
\frac{1}{(2\pi)^2 P_n^{(n)} P_z^{(p)}}
\int_0^{2\pi} \hspace{-1.5mm} d\theta \int_0^{2\pi} \hspace{-1.5mm} d\varphi \;
e^{\zI (n\theta + z\varphi)}
\nonum\\[0.5mm] && \hspace{-10mm} \times 
\det\mathcal{B}(\theta,\varphi) \int_{V_\mathrm{P}} d{\bf r}\;
\tilde{\mathcal{V}}[{\bf r},\theta,\varphi],
\label{def_E_nz_Skyrme}
\end{eqnarray}
where $\det\mathcal{B}(\theta,\varphi) \equiv
\det\mathcal{B}^{(n)}(\theta)\det\mathcal{B}^{(p)}(\varphi)$.
The Coulomb energy is evaluated using the transition proton
density, $\tilde{\rho}^{(p)}({\bf r},\theta)$.

We then define the excitation energy of the PLF composed of
$n$ neutrons and $z$ protons as
\begin{equation}
E_{n,z}^{*\,\mathrm{PLF}}(E,b) \equiv
\mathcal{E}_{n,z}^\mathrm{PLF}(E,b) - E_{n,z}^\mathrm{g.s.},
\label{E_ex_nz}
\end{equation}
where
\begin{equation}
\mathcal{E}_{n,z}^\mathrm{PLF} \equiv
\mathcal{E}_{n,z,\mathrm{kin}}^\mathrm{PLF} +
\mathcal{E}_{n,z,\mathrm{int}}^\mathrm{PLF} +
\mathcal{E}_{z,\mathrm{Coulomb}}^\mathrm{PLF}
\end{equation}
denotes the energy expectation value of the PLF.
$E_{n,z}^\mathrm{g.s.}$ is the ground state energy of
the nucleus composed of $n$ neutrons and $z$ protons.
We can also evaluate excitation energy of the TLF
in a similar way.

\subsection{Particle evaporation probabilities}

We evaluate the particle evaporation employing
a statistical model developed by I.~Dostrovsky and his
coworkers \cite{DostrovskyIII}. In this model, evaporation
of neutrons, protons, deuterons, tritons, $^3$He, and $\alpha$
particles are taken into account. An input of the model is
the excitation energy of a nucleus to be disintegrate. For more
detail explanation of the model, see Ref.~\cite{DostrovskyIII}
and references therein.

Putting as an input the excitation energy evaluated from the
TDHF wave function after collision using  Eq.~(\ref{E_ex_nz}),
we evaluate evaporation processes. Starting with the excited
fragment, all possible decay sequences reaching to the final
state at which any emissions of a particle are energetically
prohibited are considered. Each series of the evaporation
is called as the evaporation cascade. In each evaporation
cascade, the kind of emitted particles and its kinetic energy
are selected stochastically.

As an example, let us consider a case that we calculate evaporation 
processes from an excited PLF composed of $N$ neutrons and $Z$ 
protons with excitation energy of $E_{N,Z}^{*\,\mathrm{PLF}}$.
If a nucleus composed of $N'$ neutrons and $Z'$ protons is formed
at the end of an evaporation cascade, the total number of evaporated
neutrons and protons are given by $N-N'$ and $Z-Z'$, respectively.
We count the number of cases in which $n$ neutrons and $z$ protons
are evaporated until the end of an evaporation cascade among all
of the evaporation cascades examined. Then, we calculate the
evaporation probability of $n$ neutrons and $z$ protons as
\begin{equation}
P_{n,z}^\mathrm{evap.}\bigl[ E_{N,Z}^{*\,\mathrm{PLF}}(b) \bigr]=
N_{n,z}/N_\mathrm{cascade},
\end{equation}
where $N_{n,z}$ denotes the total number of processes in which
$n$ neutrons and $z$ protons are emitted until the end among
all of the evaporation cascades. $N_\mathrm{cascade}$ denotes
the total number of evaporation cascades examined. Because the
excitation energy, $E_{N,Z}^{*\,\mathrm{PLF}}$, depends on
the impact parameter, resulting evaporation probabilities,
$P_{n,z}^\mathrm{evap.}$, also depend on the impact parameter.

\subsection{Transfer cross sections with evaporation effects}

In our previous article \cite{KS_KY_MNT}, we calculated a transfer
cross section for the channel in which a PLF is composed of
$N$ neutrons and $Z$ protons by integrating the probability
$P_{N,Z}(b)$ over the impact parameter, as
\begin{equation}
\sigma_\mathrm{tr}(N,Z) = 2\pi 
\int_{b_\mathrm{min}}^\infty
b \hspace{1.5mm} P_{N,Z}(b)\, db.
\label{sigmatot_old}
\end{equation}
The minimum impact parameter of the integration 
was taken to be a border dividing fusion and binary reactions.
Here, we have assumed that both projectile and target nuclei
are spherical, so that the reaction is specified by the incident
energy $E$ and the impact parameter $b$. In practice, we first
examined the maximum impact parameter, $b_\mathrm{f}$,
in which fusion reactions take place for a given incident energy.
We then repeated reaction calculations at various impact
parameters for the region, $b>b_\mathrm{f}$, and calculated
the cross section by numerical quadrature according to
Eq.~(\ref{sigmatot_old}).

To include effects of particle evaporation into the cross section,
we simply extend the expression of the cross section by using the
evaporation probabilities obtained from the statistical calculation.
Let us denote the evaporation probability of $n$ neutrons and $z$
protons from the PLF composed of $N+n$ neutrons and $Z+z$
protons having excitation energy of $E_{N+n,Z+z}^{*\,\mathrm{PLF}}$
as $P_{n,z}^\mathrm{evap.} \bigl[ E_{N+n,Z+z}^{*\,\mathrm{PLF}} \bigr]$.
The residual nucleus after the particle evaporation is composed
of $N$ neutrons and $Z$ protons. We calculate the cross section
for the channel where the PLF is composed of $N$ neutrons
and $Z$ protons including effects of particle evaporation as
\begin{eqnarray}
\sigma_\mathrm{tr}^\mathrm{evap.}(N,Z) &=& 
2\pi \int_{b_\mathrm{min}}^\infty b \,
\sum_{n,z} P_{N+n,Z+z}(b) \nonum\\
&&\times \;P_{n,z}^\mathrm{evap.}
\bigl[ E_{N+n, Z+z}^{*\,\mathrm{PLF}}(b) \bigr]\, db.
\label{sigmatot_new}
\end{eqnarray}

\section{Results}\label{Sec:results}

\begin{figure}[t]
\centering
\includegraphics[width=8cm]{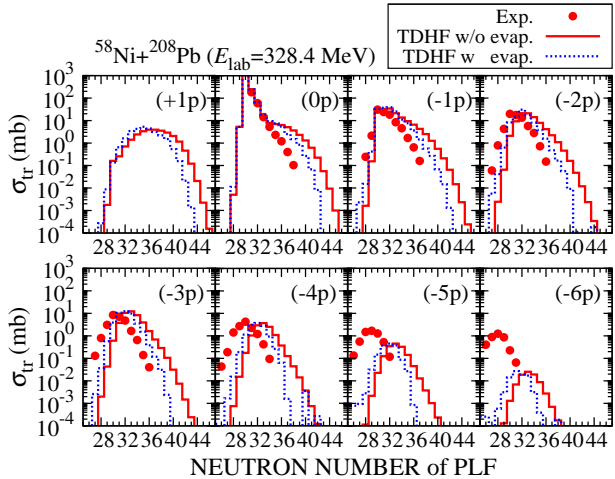}
\caption{(Color online)
Production cross sections of the projectile ($^{58}$Ni) like fragments 
in $^{58}$Ni+$^{208}$Pb reaction at $E_\mathrm{lab}=$ 328.4~MeV.
Solid (dotted) Line shows cross sections calculated by the TDHF theory
without (with) evaporation effects. Measured cross sections
\cite{Corradi(58Ni+208Pb)} are also shown by filled circles.
}
\label{fig-1}
\end{figure}

We consider $^{58}$Ni+$^{208}$Pb reaction, which was also
studied in Ref.~\cite{KS_KY_MNT}. In Fig.~\ref{fig-1}, we show cross
sections classified according to the change of the proton
number of the PLF from $^{58}$Ni, as functions of the
neutron number of the PLF. Red filled circles denote measured
cross sections \cite{Corradi(58Ni+208Pb)} and red solid
(blue dotted) lines denote results of the TDHF calculations
without (with) effects of particle evaporation. As the figure
shows, the TDHF theory describes surprisingly well the measured
cross sections when the number of transferred nucleons is small.
We note that there is no empirical parameter in our calculations,
since we employ a standard Skyrme effective interaction (SLy5
\cite{Chabanat}). As the number of transferred nucleons increases,
there appear discrepancies even when we include evaporation
effects. This fact may indicate significance of correlation effects
which are not included in the framework of the TDHF theory.
A more detail investigation will be presented in
the forthcoming paper \cite{article2}.

\section{Summary}\label{Sec:summary}

In the article, we have presented an outline of our attempt
to include effects of particle evaporation in the description
of multi-nucleon transfer processes based on the microscopic TDHF theory. 
We evaluate excitation energy of a produced
fragment in each transfer channel extending the particle number
projection technique. We calculated multi-nucleon
transfer cross sections for $^{58}$Ni+$^{208}$Pb reaction 
in the TDHF theory with and without evaporation effects, and
compared with measured cross sections. We have found that
the inclusion of evaporation effects improves the cross section
towards the measurements. However, calculations still underestimate 
measured cross sections when a number of protons are transferred. 
\vspace{5mm}

\begin{acknowledgement}
This work is supported by the Japan Society of the Promotion of
Science (JSPS) Grants-in-Aid for Scientific Research Grant Number
23340113
, and by the JSPS Grant-in-Aid for
JSPS Fellows Grant Number 25-241.
\end{acknowledgement}

%
%

\end{document}